\documentclass[sigconf]{acmart}

\usepackage{amssymb,amsfonts}   

\usepackage{subcaption}
\usepackage{colortbl}
\usepackage{algorithm}

\newsavebox{\tablebox}
\newcommand{\fitcolumn}[1]{\sbox{\tablebox}{#1}\ifdim\wd\tablebox>\columnwidth\resizebox{\columnwidth}{!}{\usebox{\tablebox}}\else\usebox{\tablebox}\fi}
\usepackage{algorithmic}

\newcommand{\red}[1]{\textcolor{black}{#1}}
\newcommand{\redHL}[1]{\textcolor{black}{#1}}

\newcommand{\blue}[1]{\textcolor{black}{#1}}

\newcommand{\ignore}[1]{}

\makeatletter\let\ACM@origbaselinestretch\baselinestretch\makeatother

\title{On the Limits of Ising Machines for Circuit-Derived SAT}

\author{
Ahmet Efe,
Abhimanyu Kumar,
Nafisa S. Prova,
H\"{u}srev~C{\i}lasun,
Ziqing Zeng,
Tahmida Islam, \\
Ruihong Yin, 
Chaohui Li,
Peter Kreye,
Chris H. Kim,
Sachin S. Sapatnekar,
Ulya R. Karpuzcu \\
\textit{University of Minnesota, Twin Cities}\\
\{efe00002, ukarpuzc\}@umn.edu
}

\copyrightyear{2026}
\acmYear{2026}
\setcopyright{cc}
\setcctype{by}
\acmConference[ICCAD '26]{IEEE/ACM International Conference on Computer-Aided Design}{November 08--12, 2026}{San Jose, CA, USA}
\acmBooktitle{IEEE/ACM International Conference on Computer-Aided Design (ICCAD '26), November 08--12, 2026, San Jose, CA, USA}
\acmISBN{979-8-4007-2873-0/2026/11}
\acmDOI{10.1145/3831252.3834256}


\begin{document}

\begin{abstract}
Emerging Ising machines are promising for solving computationally hard optimization problems, yet their limits on structured, circuit-derived satisfiability (SAT) problems remain poorly understood. Using semiprime factorization as a representative benchmark, we show that tight constraints, when mapped into optimization form, fundamentally distort the energy landscape, and that these distortions are amplified when problems are decomposed to fit limited Ising machine capacity. To address this, we propose a hybrid flow that offloads Ising-harmful structure to lightweight preprocessing while reserving the genuinely hard search for the Ising machine. We further show that generic, circuit-structure-unaware decomposition is insufficient for circuit-derived instances, and that structure-aware partitioning is essential. These findings identify constraint handling as a central obstacle, highlighting hybrid hardware-software approaches as the path forward for scaling Ising machines to real-world SAT workloads. Evaluated on fabricated 45-spin all-to-all Ising chips, our flow extends solvable problem sizes from 8-bit (94 variables) to 11-bit (190 variables) without any hardware changes.
\end{abstract}

\maketitle

\section{Introduction}
Ising machines are emerging as hardware accelerators 
for combinatorial optimization. Most performance
characterization studies use randomly generated
benchmarks which fall short of capturing the
domain-specific internal structure of real-world
problems. When such problems are mapped to an Ising
formulation, constraints become energy
penalties, but the structure itself is not translated
into the reasoning shortcuts that classical
solvers exploit. On the contrary, structure-unaware
problem mapping oftentimes hinders convergence to a
solution. In electronic design automation, for example,
satisfiability (SAT) problems derived from hardware
designs inherit gate structure, fixed signal values,
deterministic logic relations, and tightly coupled local
constraints at the gate-level, making circuit-derived
SAT instances challenging for Ising machines.
Understanding these limitations---and developing practical 
methods to overcome them---is the central goal of this 
paper.

Our central claim is that {\em tight constraints},
abundant in circuit-derived problems, pose the
primary challenge for Ising
machines. While such constraints help classical solvers 
to prune the search space, they appear in the Ising 
formulation only as penalty terms, distorting the energy 
landscape without providing any reasoning benefit. 
Problem decomposition exacerbates this effect: when 
problem instances are too large to fit on fixed-size 
Ising hardware and must be solved in pieces, incorrect 
partial assignments can propagate error across iterations, 
making convergence harder rather than easier. We investigate this effect using SAT encodings of
semiprime factorization, a particularly challenging form
of integer factorization, as a representative case study.

\begin{figure}[tp]
\centering
\includegraphics[width=1\linewidth, 
    trim={0.7cm 0.4cm 0.7cm 0.1cm}, clip]
    {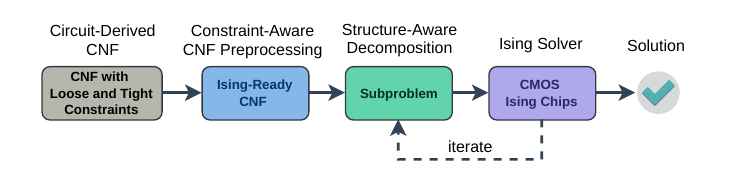}
\vspace{-0.5cm}
\caption{Overview of the proposed flow. Tight 
deterministic structure is extracted before SAT-to-QUBO 
conversion, leaving a smaller residual problem for 
decomposed Ising execution on fixed-size hardware.}
\Description{Block diagram of the proposed flow: a circuit-derived CNF passes through constraint-aware preprocessing and SAT-to-QUBO conversion, and structure-aware decomposition then feeds subproblems to the Ising solver.}
\label{fig:flow_overview} \vspace{-.6cm}
\end{figure}

Specifically, we propose a hybrid hardware--software
flow (Figure~\ref{fig:flow_overview}). We
start with the standard CNF (Conjunctive Normal Form)
representation of the problem. We combine 
constraint-aware preprocessing with structure-aware 
decomposition to maximize the Ising machine's efficiency 
in exploring the solution space. The preprocessing step 
removes structure that is cheap to resolve symbolically 
but harmful in Ising form---known signal values, 
deterministic relations, and unfavorable gate 
encodings---before the problem reaches the Ising machine, 
reducing distortion and leaving more hardware budget 
available for genuine search. The decomposition step 
preserves the domain-specific problem structure in 
forming subproblems: following the natural signal flow 
through the circuit keeps meaningful dependencies inside 
the active subproblem, whereas blindly partitioning 
isolates gate fragments that are locally consistent but 
globally misleading. The two components are 
complementary---preprocessing reduces what the hardware 
must search, and structure-aware decomposition ensures 
that what remains is presented in a form the hardware can 
use effectively.

We evaluate the proposed flow on a fabricated 45-spin 
all-to-all CMOS Ising chip. Our results show that tight 
constraints are the central source of 
difficulty---confirmed on semiprime\blue{, ATPG, and
equivalence-checking} instances \blue{as well as}
synthetic benchmarks with varying constraint
density---with the same preprocessing changing classical
solver performance by less than $3\times$ but Ising 
performance by over $150\times$, indicating a formulation 
mismatch rather than a device limitation. On the 
semiprime benchmark set, our flow doubles the largest 
solvable problem size on the same chip without any change 
to the hardware. Our main contributions are as follows:

\begin{list}{\labelitemi}{\leftmargin=0.5em}

\item We identify tight constraints, prevalent in
circuit-derived problems, as the primary challenge for
Ising machines, and show that circuit structure cannot be 
exploited effectively unless it is explicitly extracted 
before mapping.

\item We develop a constraint-aware preprocessing flow 
that removes Ising-harmful deterministic structure before 
hardware execution, reducing formulation distortion and 
increasing the fraction of the hardware budget available 
for genuine search.

\item We show that decomposition must respect 
circuit structure, and that following signal flow through 
the circuit consistently outperforms generic partitioning 
on circuit-derived instances.

\item We demonstrate that our flow can double the largest 
solvable problem size on fabricated Ising chips without 
any changes to the hardware design.

\end{list}

The remainder of this paper is organized as follows. 
Section~\ref{sec:back_and_related} reviews background 
and related work; Section~\ref{sec:limits} characterizes 
the limits of Ising machines in solving circuit-derived 
SAT; Section~\ref{sec:preprocess} presents the proposed 
preprocessing and decomposition flow; 
Section~\ref{sec:evaluation} provides the evaluation; and 
Section~\ref{sec:conclusion} concludes the paper.


\section{Background and Related Work}
\label{sec:back_and_related}

\subsection{Circuit-Derived SAT and Structured Problems}

Many practical SAT problems in electronic design automation---ranging from 
equivalence checking~\cite{kuehlmann2002robust} to bounded model 
checking~\cite{biere1999symbolic} and test pattern 
generation~\cite{larrabee1992test}---are \emph{circuit-derived}: 
a hardware design is encoded gate-by-gate into the standard Conjunctive Normal Form (CNF) via the Tseitin 
transformation~\cite{tseitin1983complexity}. The resulting CNFs 
inherit non-uniform variable interactions, locality, and modular 
organization from the underlying 
circuit~\cite{ansotegui2019community}, distinguishing them sharply 
from the uniformly generated random problem instances commonly used to 
evaluate Ising-based SAT 
solvers~\cite{mitchell1992hard,cheeseman1991really}. Crucially, 
circuit-derived CNFs have specific structure.
\blue{Unlike} Ising machines, classical solvers
can identify this structure to prune the search space. 

We use semiprime factorization as a representative circuit-derived 
benchmark. We encode the corresponding multiplication circuit with outputs fixed to a known 
product 
into CNF using the Tseitin transformation, which renders a SAT instance
whose satisfying assignment recovers the input 
factors~\cite{mosca2022factoring}. 
\blue{Semiprime} CNFs cover a wide range of arithmetic structures such as half adders, full adders, carry chains, or AND 
gates. We control problem hardness by scaling the bit-width.
Prior work has used semiprime factorization 
for embedding studies or for performance characterization
on quantum 
hardware~\cite{ding2024factorization,jun2023hubo,jiang2018quantum}.
We instead investigate the impact of circuit-derived structures on Ising machine performance. 


\subsection{Ising Machines, QUBO, and Decomposition}

Ising machines reach equilibrium at configurations {\bf s} that minimize the Hamiltonian
\begin{equation}
H(\mathbf{s}) = \redHL{-} \sum_{i<j} J_{ij}\, s_i s_j \redHL{-} \sum_i h_i\, s_i,
\end{equation}
where $s_i \in \{-1,+1\}$; $J_{ij} \in \mathbb{R}$; $h_{i} \in \mathbb{R}$~\cite{lucas2014ising}. 
Implementations span CMOS oscillator 
arrays~\cite{cilasun2024sat,ahmed2021probabilistic}, quantum 
annealers~\cite{johnson2011quantum}, and photonic 
systems~\cite{inagaki2016coherent}. Through the substitution 
$x_i = (1{+}s_i)/2$, the Ising model is equivalent to Quadratic 
Unconstrained Binary Optimization 
(QUBO)~\cite{glover2018tutorial}. Under the Chancellor 
formulation~\cite{chancellor2016direct}, a problem with $n$ 
variables and $m$ clauses maps to $n{+}m$ spins for 3SAT, 
while 1SAT and 2SAT require no extra spins beyond problem variables. As incomplete 
solvers~\cite{gomes2008satisfiability}, Ising machines can find 
satisfying assignments but cannot certify unsatisfiability. When 
a problem instance exceeds hardware capacity, it must be partitioned and 
solved iteratively~\cite{boost2017partitioning}. 
Variable Interaction Graphs visualize constraints as variable-to-variable interactions (connections). 
In forming subproblems, Breadth First Search (BFS) based variable 
selection on the Variable Interaction Graph has been shown effective
for random SAT~\cite{cilasun2024sat}\blue{\cite{efe2026satic}}. We revisit this
for circuit-derived problems.

\subsection{Related Work}

\noindent {\bf Ising-based SAT Solvers:}
Prior work has focused primarily on random or loosely structured 
problem instances. C{\i}lasun et al.~\cite{cilasun2024sat} demonstrates 3SAT on 
a fabricated 
CMOS Ising chip using multiple QUBO 
formulations and decomposition strategies, but 
evaluates only
random SATLIB benchmarks.
HyQSAT~\cite{tan2023hyqsat} introduces a hybrid solver where Ising hardware assists a 
classical (CDCL) solver rather than serving as the primary solver. 
\blue{Our flow reverses these roles: the Ising machine is the
primary optimization engine, while software makes the problem
Ising-ready before hardware execution.}
Ding
et al.~\cite{ding2024factorization} 
represents
one of the few efforts 
solving
circuit-derived SAT using \blue{D-Wave} 
quantum annealers, but does not analyze
why circuit structure makes the 
problem harder for Ising machines.

\noindent{\bf SAT-to-Ising (QUBO) Mapping:}
Chancellor et al.~\cite{chancellor2016direct} and N{\"u}{\ss}lein 
et al.~\cite{nusslein2023solving} propose direct $k$SAT-to-QUBO 
mappings with different overhead (in terms of extra spins beyond problem variables). 
Sikhakollu et 
al.~\cite{sikhakollu2024circuit} introduces circuit mappings 
for oscillator based Ising machines. Jagielski et 
al.~\cite{jagielski2023fpim} uses logic synthesis to map SAT onto 
an FPGA-based Ising machine. More compact (higher-order) Ising formulations 
have also been explored
~\cite{bybee2023higher,zhang2024skisat}. These efforts 
improve the mapping itself but do not address the structural 
mismatch between circuit-derived CNFs and the QUBO energy landscape 
that we investigate in this work.

\blue{We focus on} the sensitivity of SAT-to-QUBO
conversion and hardware-driven decomposition to
the internal structure of 
circuit-derived CNFs.

\section{Ising Machine Limits on Circuit-Derived SAT}
\label{sec:limits}
Previous studies have demonstrated how SAT can be mapped to Ising machines and solved using decomposition~\cite{cilasun2024sat,chancellor2016direct}. What remains less understood is why circuit-derived SAT instances are substantially harder than random problems. We show that the central mechanism is \emph{tight constraints}---highly restrictive relations that substantially reduce local search freedom---which arise abundantly in circuit-derived CNFs and interact unfavorably with both SAT-to-QUBO conversion and hardware-driven decomposition.

\subsection{Tight Constraints in Subproblems}
\label{sec:limits_qubo}

Under standard SAT-to-QUBO translation the optimization objective becomes
\begin{equation}
E(x) = \sum_{c \in \mathcal{C}} \lambda_c \, 
\phi_c(x),
\label{eq:penalty}
\end{equation}
where $x \in \{0,1\}^n$ captures problem variables, $\lambda_c > 0$ is the penalty weight of clause $c$, and $\phi_c(x)=0$ if a given clause $c$ is satisfied and $\phi_c(x)>0$ otherwise. Solutions \blue{satisfy} $E(x)=0$.

Under this formulation, tight constraints appear only as penalty terms: they do not provide the propagation and pruning benefits that classical solvers derive from them, and they can still be violated during optimization. This applies even to constraints imposed by constant logic values---because Ising machines are probabilistic solvers, even a logically determined value may be violated in the returned solution. Worse, penalty terms can suppress each other in destructive ways. For example, consider a problem with a mix of 1SAT and 2SAT clauses over variables $a$ and $b$. A unit (1SAT) clause \blue{$(\neg a)$} imposes the tight constraint \blue{$a{=}0$} through a linear term $a$ in the energy function. A 2SAT clause \blue{$(a \lor b)$} over the same variable, however, introduces quadratic QUBO coefficients such as $-a - b + ab$ that can eliminate the 1SAT term, hence the only encoding of the underlying tight constraint. Ising solvers in return can consider assignments with \blue{$a{=}1$} as solutions, which clearly violate a tight problem constraint. In circuit-derived CNFs, fixed product bits propagate through half adders, full adders, and carry chains, creating tightly coupled local relations that concentrate penalty structure across the formula without any reasoning benefit to the hardware.

\subsection{Tight Constraints Across Subproblems}
\label{sec:limits_decomp}

Decomposition exacerbates the situation. We need to form subproblems if a problem instance is too large for the hardware. We refer to the subset of variables included in a given subproblem as an {\em active} set $X_A$. The rest of the variables form a {\em frozen} set $X_F$. Each subproblem is solved under a specific assignment $\hat{x}_F$ to frozen variables to minimize the conditioned energy
\begin{equation}
E_A(x_A \mid \hat{x}_F).
\label{eq:conditioned}
\end{equation}
The conditioned feasible set is
\begin{equation}
S(\hat{x}_F)=\{x_A \mid F(x_A,\hat{x}_F)=1\},
\label{eq:residual}
\end{equation}
where $F$ denotes the original CNF. If $\hat{x}_F$ is globally consistent, $S(\hat{x}_F)$ still contains valid \red{completions} \blue{(assignments to $X_A$ that satisfy $F$ under $\hat{x}_F$)}. If $\hat{x}_F$ is wrong, however, $S(\hat{x}_F)$ can shrink sharply or become empty, while the Ising machine continues to optimize the conditioned energy regardless.

If problem decomposition is not aware of tight constraints, variables encoding tight constraints can be frozen with incorrect logic assignments. The resulting subproblem therefore becomes more likely to assign infeasible values to its active set, which in turn becomes the frozen set for the remaining subproblems solved in subsequent iterations. As a result, the Ising machine may still find solutions to individual subproblems, however, not to the original problem due to conflicting partial solutions. Subproblems in circuit-derived SAT share variables by construction (gates share variables, carry chains connect logic layers, ...), so a wrong frozen assignment at one stage constrains the next. The solver refines a trajectory that is locally plausible but globally wrong, and correcting that trajectory becomes \blue{harder} as incorrect context propagates across iterations.

\subsection{Why Subproblem Size Matters}
\label{sec:limits_rho}

The previous two subsections show that tight constraints are harmful both when they remain inside the active QUBO and when they are left outside as frozen context. A third consequence follows directly: they consume hardware budget. On fixed-size hardware, every spin spent on an already-determined variable, an ancillary spin introduced by clause-to-QUBO conversion, or other structural overhead is a spin not available for genuine search. This is especially important for circuit-derived CNFs, where logically equivalent formulations can differ substantially in the overhead incurred by extra spins: a 3SAT clause may require ancillary spins, whereas a reduced 2SAT clause does not, and an unfavorable gate encoding may consume more hardware than an equivalent Ising-friendlier form.

To quantify this effect, we define the {\em active fraction} per subproblem as
\begin{equation}
\rho = \frac{|X_A|}{n},
\label{eq:active_fraction}
\end{equation}
where $|X_A|$ is the number of active (non-frozen) variables and $n$ is the total variable count of the original SAT instance. \blue{For a given instance, a} low $\rho$ means that most of the hardware resources, and hence, the time spent in exploring the solution space, go to extra spins not having a direct correspondence to problem variables, hindering convergence.

This makes preprocessing valuable for reasons beyond ordinary SAT simplification. Eliminating known values removes already-determined variables from the hardware problem. Clause cleaning and propagation can reduce 3SAT clauses to 2SAT, thereby removing ancillary overhead. Gate-aware rewriting can replace logically equivalent but hardware-expensive encodings with more compact ones. Together, these steps reduce compiled problem size while increasing $\rho$, allowing more of the fixed hardware budget to be spent on the genuinely hard part of the search.

\subsection{Tight Constraints Beyond Circuit Structure}
\label{sec:limits_backbone}

The analysis above is grounded in circuit-derived structure, but the underlying cause should be observable more broadly. Backbone variables---those taking the same value in every satisfying assignment---provide a controllable proxy for tight-constraint pressure. As backbone density $\beta = |B|/n$ increases, the formula contains more logically fixed structure and fewer valid solutions, making the search harder for a probabilistic solver. If tight constraints are themselves the source of difficulty---rather than an artifact of semiprime structure---then increasing $\beta$ should degrade Ising performance even on synthetic random instances. Section~\ref{sec:eval_backbone} confirms this prediction on controlled-backbone SATLIB benchmarks across both our fabricated chip and a software QUBO baseline.

Together, these observations lead to a single design rule: tight deterministic structure should be removed before QUBO conversion when it can be resolved cheaply. The goal is not conventional SAT preprocessing, but preprocessing \emph{for Ising execution}---explicitly
extracting the structure that classical solvers exploit implicitly, before it survives as harmful penalty terms, incorrect frozen context, or wasted hardware budget. Section~\ref{sec:preprocess} develops this into a concrete constraint-aware preprocessing and
decomposition flow.

\section{Constraint-Aware Preprocessing \& Structure-Aware Decomposition}
\label{sec:preprocess}
The analysis in Section~\ref{sec:limits} leads to a concrete hybrid flow. First, lightweight symbolic preprocessing removes tight deterministic structure that would otherwise survive as harmful penalty terms inside the active QUBO or as harmful frozen context outside it. The residual problem is then solved through iterative decomposition on fixed-size hardware using a subproblem selection strategy that respects circuit structure.

Our preprocessing targets three sources of Ising-harmful structure: gate encodings whose CNF form is unfavorable for QUBO mapping, known values induced by fixed outputs, and deterministic two-variable relations that symbolic solvers exploit naturally but Ising machines cannot. Algorithm~\ref{alg:preprocess} summarizes the preprocessing flow. These transformations are applied once, on the globally consistent CNF, before decomposition begins.

\begin{algorithm}[t]
\footnotesize
\caption{Constraint-Aware Preprocessing}
\label{alg:preprocess}
\begin{algorithmic}[1]
\REQUIRE CNF formula $\mathcal{F}$
\ENSURE Residual CNF $\mathcal{F}_{\mathrm{res}}$, 
        fixed-value map $\mathcal{A}$, 
        replacement map $\mathcal{R}$
\STATE Initialize $\mathcal{A} \gets \emptyset$, 
       $\mathcal{R} \gets \emptyset$
\REPEAT
    \STATE Rewrite unfavorable gate encodings into Ising-ready CNF forms
    \STATE Propagate unit clauses and record fixed 
           values in $\mathcal{A}$
    \STATE Detect deterministic binary relations and 
           update $\mathcal{R}$
    \STATE Apply replacements from $\mathcal{R}$
           to $\mathcal{F}$ and propagate fixed
           values transitively through $\mathcal{R}$
    \STATE Clean clauses: remove satisfied clauses, 
           falsified literals, duplicates, and 
           tautologies
\UNTIL{a full sweep yields no new fixed values, 
       replacements, or clause-width reductions}
\RETURN $\mathcal{F}_{\mathrm{res}}=\mathcal{F}$, 
       $\mathcal{A}$, $\mathcal{R}$
\end{algorithmic}
\end{algorithm}

\subsection{Constraint-Aware Preprocessing}

\noindent{\bf Gate-Aware CNF Rewriting:}
A circuit gate can be represented by multiple logically equivalent CNF encodings, but these encodings are not equivalent for Ising execution. Symbolic SAT solvers are largely insensitive to which encoding is used; for Ising-oriented compilation, however, the encoding directly affects the number of ancillary variables introduced during SAT-to-QUBO conversion and therefore the hardware footprint of each gate.

Consider the OR-gate relation $c \leftrightarrow (a \lor b)$. A direct encoding yields a \blue{3SAT} clause for each invalid assignment, resulting in four clauses that require 7 physical spins under the Chancellor formulation~\cite{chancellor2016direct}. Alternatively, the same gate can be expressed via a Tseitin-style transformation as
\begin{equation}
(\neg c \lor a \lor b)
\land (\neg a \lor c)
\land (\neg b \lor c),
\label{eq:or_opt2}
\end{equation}
which requires only 4 spins, since only the 3SAT clause introduces an ancillary variable---a \blue{43\%} reduction in per-gate spin cost.

Gate detection is purely syntactic: clauses sharing a common support set are grouped, and their literal-polarity patterns are matched against known gate signatures. This rewriting applies to AND, OR, NAND, and NOR gates; XOR and XNOR gates are already expressed in forms that cannot be compacted further. Gate-aware rewriting leaves the variable count unchanged but directly increases $\rho$ by reducing the ancillary overhead consumed per gate. On benchmarks where AND- and OR-type gates dominate---as in the semiprime multiplier circuits studied here---this step meaningfully reduces the compiled QUBO size before any variables are eliminated.

\noindent{\bf Known-Value Elimination:}
Circuit-derived SAT instances frequently contain assignments that are determined a priori. In formal verification, miter outputs are constrained to represent inequivalence; in ATPG, target signals are assigned specific logic values; and in semiprime factorization, product bits are fixed by the problem definition. While symbolic SAT solvers immediately resolve the resulting unit clauses via value propagation, Ising machines treat a known value such as $a=1$ as a penalty term $\lambda_a(1-a)$ rather than a resolved fact. During problem decomposition, such variables may inadvertently freeze to incorrect states, propagating conflicting assignments through dependent logic.

We eliminate known values before QUBO conversion: whenever a variable is fixed by a unit clause, we record its assignment in $\mathcal{A}$ (Algorithm~\ref{alg:preprocess}), remove all satisfied clauses, and shorten clauses containing the opposite literal. This procedure is applied recursively to capture the full transitive closure of logical implications. For instance, fixing an OR-gate output to 0 forces both inputs to 0, potentially triggering further unit clauses. Furthermore, this propagation can reduce 3SAT clauses to 2SAT, thereby removing ancillary overhead because 2SAT clauses require no ancillary spins under the Chancellor formulation. These assignments are maintained as external invariants rather than embedded into the QUBO.

\begin{figure}[t]
\centering
\includegraphics[width=\linewidth]
    {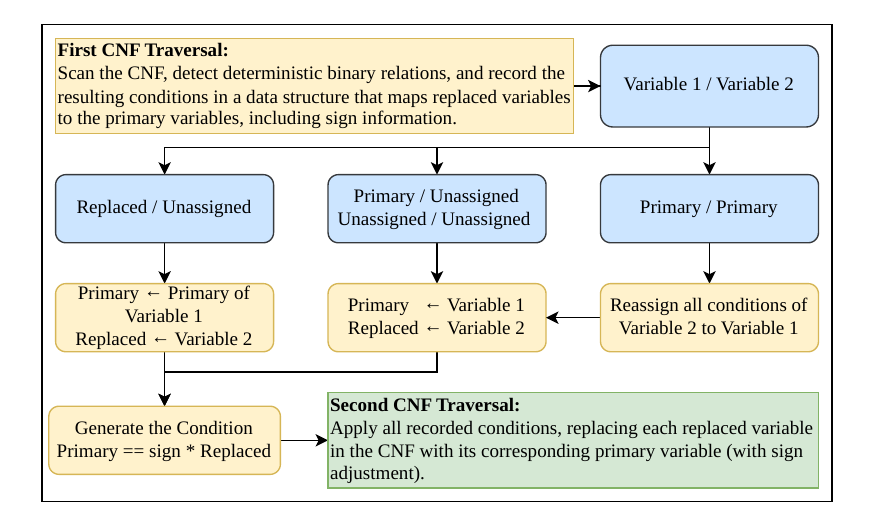}
\vspace{-0.9cm}
\caption{Two-pass decision logic for deterministic binary relation elimination. Variables are classified as {\em primary}, {\em replaced}, or {\em unassigned}; the action taken depends on the joint classification of both variables in the detected relation.}
\Description{Flowchart of the two-pass classification used to eliminate deterministic binary relations, showing the action taken for primary, replaced, and unassigned variable pairs.}
\label{fig:2SAT_conditioning_decision}
\vspace{-0.5cm}
\end{figure}

\noindent{\bf Deterministic Binary Relation Elimination:}
After known-value elimination, circuit-derived CNFs often expose deterministic relations between pairs of variables. The most common cases are buffer and inverter patterns,
\begin{equation}
(a \lor \neg b)\land(\neg a \lor b) 
\;\Longrightarrow\; a=b,
\end{equation}
and
\begin{equation}
(a \lor b)\land(\neg a \lor \neg b) 
\;\Longrightarrow\; a=\neg b.
\end{equation}
These relations carry no genuine search freedom: once one variable is known, the other is fully determined. After SAT-to-QUBO conversion, they survive as penalty structure that consumes spin budget and distorts the energy landscape without \blue{aiding} the search.

We excise \emph{both} clauses from the CNF and represent the underlying relation as explicit, external information maintained alongside the QUBO formulation. One variable is designated as the {\em primary}, while its dependent counterpart---{\em replaced}---is substituted with its implied value throughout the formula. Upon the completion of each Ising solver iteration, these recorded relations reconstruct the values of all substituted variables from their respective primaries, ensuring logical consistency without exhausting physical spin resources. This approach augments the natively unconstrained QUBO model with \blue{external invariants} enforced outside the energy landscape.

A naive implementation would necessitate re-traversing and re-writing the CNF upon the detection of each relation, resulting in prohibitive computational overhead. Furthermore, chained dependencies---such as $a=b$ followed by $b=\neg c$---complicate the reduction process, risking the loss of logical transitive closure. We therefore implement a two-pass algorithm that executes all substitutions in exactly two CNF traversals, deferring all modifications to the final stage (Figure~\ref{fig:2SAT_conditioning_decision}).

During the initial pass, the CNF is scanned to aggregate all relations into a map, categorizing variables into three disjoint sets: {\em primary} (retained in the CNF), {\em replaced} (to be substituted), and {\em unassigned}. Upon identifying a new relation, the classification of its constituent variables dictates the update logic: initiating a new {\em primary}, extending an existing dependency chain, or merging two distinct {\em primary} sets. In the second pass, all recorded substitutions are applied concurrently.

Substitutions are applied iteratively because eliminating one relation can expose others. In circuit-derived CNFs, such cascades are frequent across buffers, inverters, and carry-related signals. Each iteration therefore expands the set of variables confirmed to carry no genuine search freedom, \blue{removing} as much Ising-harmful structure as possible before the problem is converted to QUBO.

\noindent{\bf Replacement Propagation and Clause Cleaning:}
After known-value elimination and binary substitution, we maintain the resulting dependencies transitively. For example,
\begin{equation}
a=\neg b,\quad b=\neg c,\quad c=1
\;\Longrightarrow\;
a=\redHL{1},\quad b=\redHL{0},\quad c=1.
\end{equation}
This \emph{replacement propagation} keeps the external side-information list consistent and removes variables whose values are already implied by earlier substitutions.

The more important step is \emph{clause cleaning}. Once substitutions are applied, clauses may become satisfied, shortened, duplicated, or tautological. Such redundancy has limited effect on symbolic SAT search but is directly costly for Ising compilation: each superfluous clause potentially increases QUBO dimensionality, ancillary variable count, and physical hardware footprint. We therefore perform a comprehensive cleaning pass after each substitution cycle---removing satisfied clauses, eliminating falsified literals, and merging duplicate or tautological terms. This normalization ensures that hardware resources are dedicated to exploring the search space rather than representing logical redundancies, enhancing $\rho$.

The preprocessing step terminates upon reaching a fixed-point: when a complete traversal yields no additional variable assignments, replacement relations, or clause-width reductions. At this juncture, any further simplification would necessitate heuristic search rather than deterministic propagation; consequently, the residual problem is dispatched to the decomposition stage.

\blue{Table~\ref{tab:ising_specific} summarizes the Ising-specific role of each transformation.}

\begin{table}[t]
\centering
\footnotesize
{
\caption{\blue{Ising-specific role of each preprocessing step.}}
\label{tab:ising_specific}
\setlength{\tabcolsep}{3pt}
\fitcolumn{
\begin{tabular}{@{}p{0.22\columnwidth}p{0.745\columnwidth}@{}}
\toprule
\textbf{Transformation} & \textbf{Role in the Ising flow} \\
\midrule
Gate-aware rewriting & CDCL operates on the flat CNF without knowledge of the underlying circuit; for Ising, the chosen encoding directly fixes each gate's ancillary-spin cost after SAT-to-QUBO conversion. \\
Known-value elimination & Known values are added as unit clauses to guide classical solvers; for Ising they must instead be propagated and excised from the CNF, surviving only as external invariants so the QUBO shrinks. \\
Binary-relation elimination & Both clauses are excised and one variable substituted throughout; the relation is carried as an external invariant, making the problem hybrid: part residual QUBO, part constraints enforced outside the energy landscape. \\
Replacement propagation & Propagating substitutions and fixed values transitively keeps the invariant set consistent and shrinks clause widths (3SAT to 2SAT), cutting ancillary-spin demand and freeing spin budget for genuine search. \\
Clause cleaning & Deleting satisfied, duplicate, and tautological clauses removes redundant ancillary spins and couplings, easing both the spin-budget and coupling-coefficient demands on the hardware and directly raising $\rho$. \\
\bottomrule
\end{tabular}
}
}
\end{table}

\subsection{Structure-Aware Decomposition}
\label{sec:decomposition}

Following preprocessing, the residual problem typically exceeds the physical capacity of fixed-size Ising hardware, which necessitates an iterative decomposition strategy. In each iteration, a new subproblem is formed by selecting a subset of variables as {\em active} variables, while the remaining variables are frozen at their instantaneous best-known assignments. Upon receipt of a candidate partial solution from the hardware, the global state is updated with the active variable assignments, and the subsequent subproblem is generated following the same methodology. The Ising solver thereby navigates the high-dimensional energy landscape.

A pivotal design consideration is whether to apply preprocessing within the iterative decomposition loop; our results demonstrate that doing so is harmful rather than beneficial. Once the problem is decomposed, frozen variables may hold incorrect assignments, and the conditioned subproblem is no longer a pure SAT instance: it effectively behaves as a MaxSAT formulation, where the Ising machine minimizes local clause violations to navigate toward a global solution. While this MaxSAT behavior is beneficial---utilizing unsatisfied clauses as a ``restorative signal'' to correct erroneous assignments---symbolic simplification in this state is detrimental. Such reduction removes or distorts precisely the clauses required to penalize incorrect frozen assignments, narrowing the energy landscape in ways that reinforce incorrect frozen context rather than correcting it. Consequently, to preserve the Ising machine's ability to recover from errors, we keep preprocessing as a static pass on the globally consistent CNF before any variables are frozen.

\begin{figure}[tbp]
\centering
\includegraphics[width=0.7\linewidth]
    {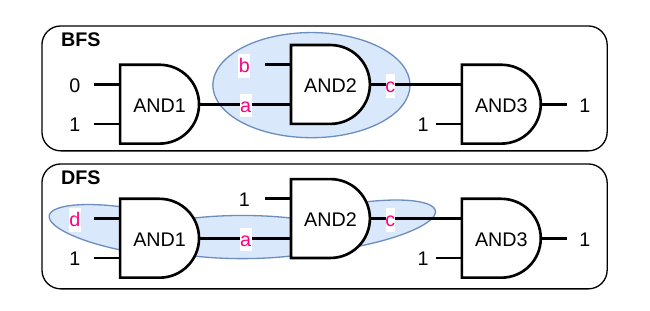}
\vspace{-0.4cm}
\caption{BFS vs.\ DFS subproblem selection on a circuit-derived CNF. BFS isolates a single gate, freezing fan-in and fan-out logic outside the chip boundary. DFS follows the signal-flow path, preserving a longer input--output chain inside the active subproblem.}
\Description{Two schematic chains of AND gates comparing BFS, which activates a single gate, with DFS, which activates a signal-flow path across several gates.}
\label{fig:bfs_vs_dfs}
\vspace{-0.5cm}
\end{figure}

Within a fixed decomposition framework, the efficiency of the solver depends on the selection strategy for the active variable set. Energy-based heuristics~\cite{boost2017partitioning} prioritize variables with high individual energy gradients, yet often fail to capture sufficient clause-level interactions, resulting in a sparse QUBO formulation that limits the solver's ability to navigate the search space. Graph-traversal-based selection addresses this by growing the active set through clause-level adjacency, ensuring that subproblems maintain logical locality by construction. BFS-based traversal has been shown effective for random SAT instances~\cite{cilasun2024sat}\blue{\cite{efe2026satic}}, where uniform connectivity allows local clusters to serve as representative subproblems.

Circuit-derived CNFs are different. In any Tseitin-encoded circuit, each gate produces a small, tightly connected clause cluster whose size is bounded by the gate's fan-in. BFS expands by adjacency within such a cluster before crossing to the next, and therefore tends to collect the inputs and output of a single gate into the active set while freezing the surrounding fan-in and fan-out logic. Since gate relations are already tightly determined, the resulting subproblem contains little genuine search freedom. Worse, if the frozen fan-in or fan-out values are incorrect---common early in search---wrong context propagates into subsequent iterations and can slow or prevent convergence. This is not a property of any particular circuit family but of the Tseitin encoding itself: any gate-by-gate CNF translation produces the same tightly clustered clause structure that BFS fails to escape.

We introduce DFS-based subproblem selection as a structure-aware alternative. Unlike neighborhood-centric traversals, DFS crosses gate boundaries by construction, following longer dependency chains through the circuit and selecting variables along primary signal-flow paths. This keeps meaningful input--output dependencies inside the active subproblem and gives the solver genuine search freedom across multiple gates in sequence. As illustrated in Figure~\ref{fig:bfs_vs_dfs}, BFS tends to isolate a single gate---such as AND2---while freezing its immediate fan-in and fan-out. This isolation frequently biases the solver toward local minima (e.g., $a{=}0$, $c{=}0$) that satisfy local constraints but violate the global \blue{constraint}. In contrast, DFS extracts a consistent logic trajectory \blue{($d{=}1$, $a{=}1$, $c{=}1$)}, ensuring that each subproblem captures a valid functional slice of the circuit that aligns with the global solution.

{\setlength{\abovedisplayskip}{3pt plus 1pt}\setlength{\belowdisplayskip}{3pt plus 1pt}\setlength{\abovedisplayshortskip}{1pt}\setlength{\belowdisplayshortskip}{1pt}The contrast between the two traversals can be quantified with a simple model. In a Tseitin-encoded circuit of two-input gates, each gate couples three variables in the variable-interaction graph. A traversal selects an active set $X_A$ of $M$ variables and freezes the rest (Section~\ref{sec:limits_decomp}). Among the active variables, let $K$ count those whose values are fully determined by the frozen context---e.g., a gate output both of whose inputs lie in $X_F$. Such variables carry no search freedom, so the subproblem's effectively searchable subspace $S$ satisfies
\begin{equation}
|S| \le 2^{\,M-K}.
\end{equation}
BFS grows a local gate cluster; in fan-in-tree-like regions, roughly half of the active variables have both inputs frozen, so $K \approx M/2$ and $|S_{\mathrm{BFS}}| \le 2^{M/2}$. DFS follows a signal-flow chain in which nearly every active variable keeps an active neighbor, leaving only the chain endpoint exposed to frozen context, so $K \approx 1$ and $|S_{\mathrm{DFS}}| \le 2^{M-1}$. Under this model,
\begin{equation}
\frac{|S_{\mathrm{DFS}}|}{|S_{\mathrm{BFS}}|} \approx 2^{\,M/2-1},
\end{equation}
which grows exponentially with subproblem size: with $M \approx 20$--$26$ active problem variables per subproblem on our 45-spin chip, DFS has the potential for a $512\times$--$4096\times$ larger searchable subspace.\par}

DFS is therefore the default strategy in our flow, while BFS is retained as a comparison baseline. Section~\ref{sec:eval_decomp} shows that DFS consistently outperforms BFS across all instance sizes and preprocessing configurations. The advantage is most pronounced on larger instances, where each subproblem \blue{covers} only a small fraction of the circuit.

\section{Evaluation}
\label{sec:evaluation}
We evaluate the proposed flow on a fabricated 45-spin all-to-all CMOS Ising chip and report results using D-Wave Tabu~\cite{dwave-tabu} as a hardware-independent probabilistic QUBO baseline.

\subsection{Experimental Setup}
\begin{figure}[t]
\centering
\includegraphics[width=0.4\linewidth]
    {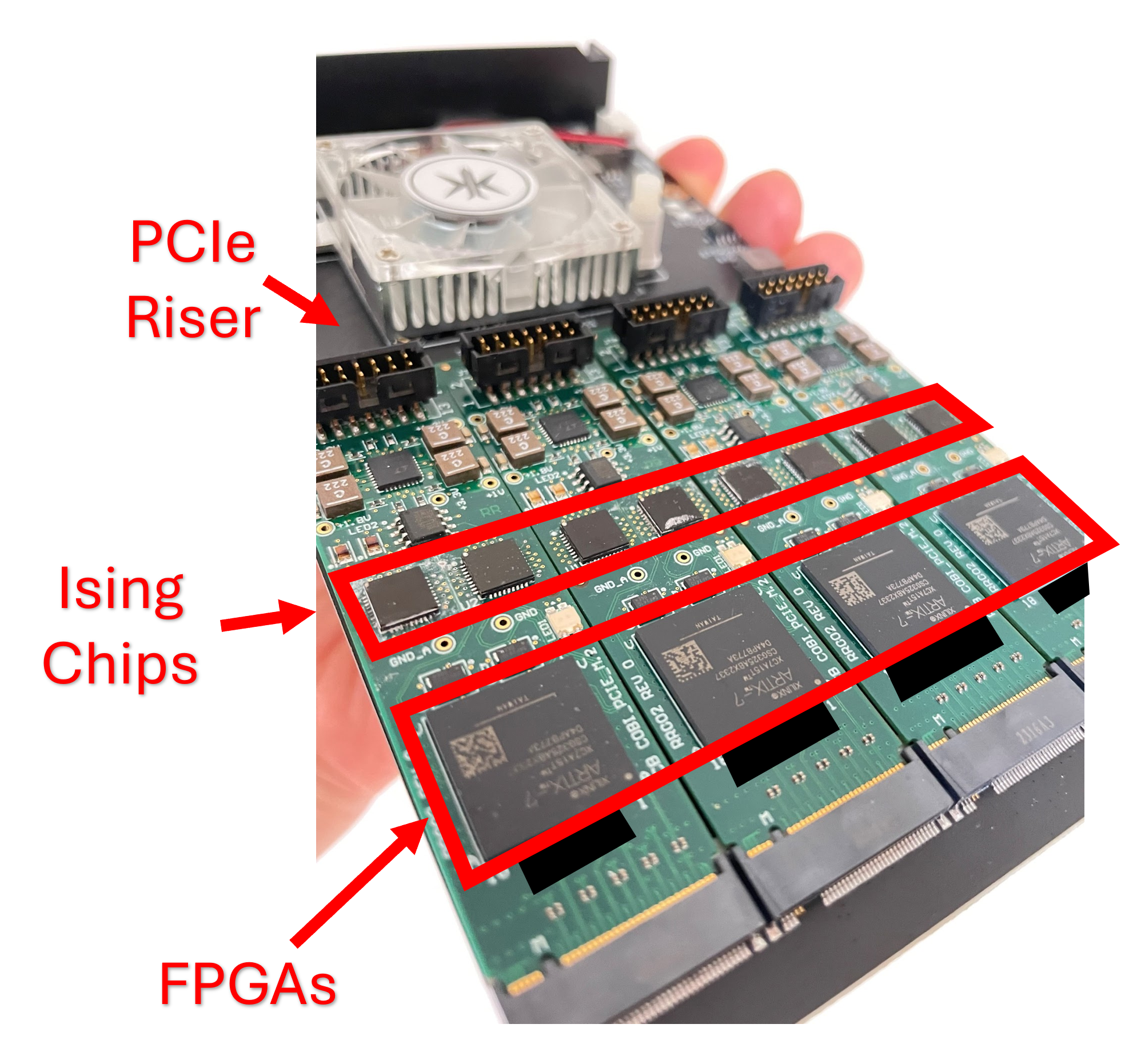}
\vspace{-0.4cm}
\caption{
45-spin all-to-all CMOS Ising testbed.
}
\vspace{-0.1cm}
\Description{Photograph of the 45-spin all-to-all CMOS Ising chip testbed mounted on its PCIe board.}
\label{fig:cobidualres}
\end{figure}

Our experiments are conducted on a fabricated platform consisting of eight all-to-all CMOS Ising chips, each with 45 usable spins \blue{(same chip family as~\cite{cilasun2025cobi}; }Figure~\ref{fig:cobidualres}).

\noindent\textbf{Metrics:}
Our primary performance metric is \emph{Time to Solution} (TTS), defined as the expected number of iterations required to solve a problem instance with 95\% success probability~\cite{ronnow2014defining}. TTS is reported only for groups in which all instances are solvable. For partially solved groups, we instead report the {\em solve rate}, {the fraction of instances solved within the iteration budget}.

\noindent\textbf{Benchmarks:}
We use \blue{three} benchmark families. SATLIB's controlled-backbone 3SAT set~\cite{hoos2000satlib} allows backbone density $\beta$ (Section~\ref{sec:limits_backbone})
to vary while holding other parameters approximately constant. A set of 79 semiprime factorization instances~\cite{quicc_sat_datasets}, encoded from schoolbook multiplier circuits with known outputs and unknown inputs, serves as the primary circuit-derived benchmark. Table~\ref{tab:instance_summary} summarizes CNF parameters and gate composition. AND- and OR-type gates admit the Ising-ready encoding described in Section~\ref{sec:preprocess}; XOR and XNOR gates cannot be rewritten in the same way. \blue{A third family, ATPG and equivalence-checking instances, is summarized in Section~\ref{sec:eval_e2e} and detailed in the supplementary material.}

\begin{table}[t]
\captionsetup{skip=2pt}
\centering
\small
\setlength{\tabcolsep}{3.5pt}
\caption{Semiprime \blue{benchmarks by}
output precision (\#Bits).}
\vspace{-1pt}
\fitcolumn{
\begin{tabular}{cccccccccc}
\toprule
& & \multicolumn{3}{c}{\textbf{CNF Parameters}} 
& \multicolumn{5}{c}{\textbf{Gate Counts}} \\
\cmidrule(lr){3-5} \cmidrule(lr){6-10}
\textbf{\#Bits} & \textbf{\#Inst.} & $k$ & $n$ & $m$ 
& AND & OR & XOR & XNOR & Total \\
\midrule
4  & 1  & 3 & 21  & 72  & 9  & 4  & 4  & 0  & 17  \\
5  & 3  & 3 & 37  & 133 & 14 & 9  & 7  & 4  & 34  \\
7  & 4  & 3 & 76  & 283 & 29 & 15 & 20 & 10 & 74  \\
8  & 11 & 3 & 94  & 352 & 35 & 16 & 26 & 18 & 95  \\
10 & 20 & 3 & 162 & 618 & 64 & 28 & 49 & 22 & 163 \\
11 & 40 & 3 & 190 & 727 & 75 & 33 & 58 & 26 & 192 \\
\bottomrule
\end{tabular}
}
\label{tab:instance_summary}
\end{table}

\noindent\textbf{Classical QUBO Baseline:}
D-Wave Tabu is configured with a 5\,ms solver timeout and a subproblem size of 45, matching the spin capacity of the fabricated Ising chips.

\noindent\textbf{Classical SAT Reference:}
We compare against Glucose and MiniSat (via PySAT~\cite{pysat}) on selected benchmarks to quantify the gap between symbolic and Ising execution. CDCL energy is measured using Intel RAPL via pyRAPL; Ising chip energy is computed from measured chip execution time at 24\,mW average Ising chip power.

\subsection{Tight Constraints Under Decomposition}
\label{sec:eval_backbone}

To quantify the impact of backbone density $\beta$ (Section~\ref{sec:limits_backbone}) on solver performance, we deploy SATLIB's controlled-backbone {size (CBS)} 3SAT benchmarks. We consider the Ising solver as well as the baseline QUBO solver (\blue{D-Wave} Tabu).

We use {CBS} instances with $n=100$ variables, clause counts $m\in\{403,411,418,423,429,435,441,449\}$, and a backbone density range of $\beta\in\{10,30,50,70,90\}\%$. All instances exceed the 45-spin hardware budget after SAT-to-QUBO conversion, so both solvers are evaluated under decomposed execution. For each $(m,\blue{\beta})$ bucket, we evaluate the first 100 instances with 100 independent repeats per solver and a per-repeat cap of 5{,}000 iterations.

\begin{figure}[t]
  \centering
  \begin{subfigure}[b]{0.49\linewidth}
    \centering
    \includegraphics[width=\linewidth]
        {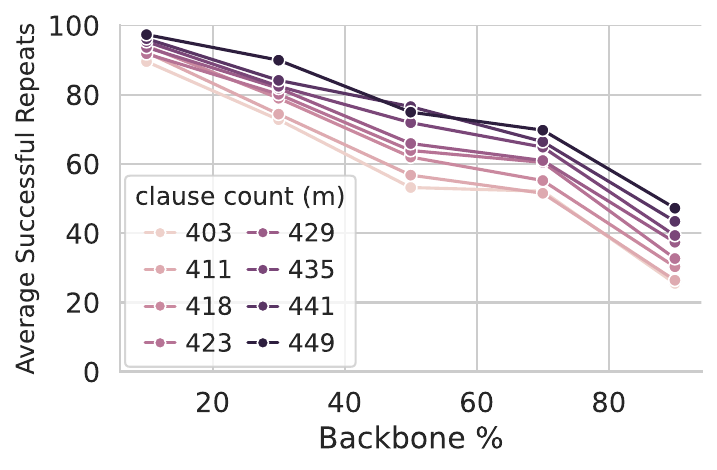}
    \caption{D-Wave Tabu}
    \label{fig:backbone_tabu_repeat}
  \end{subfigure}\hfill
  \begin{subfigure}[b]{0.49\linewidth}
    \centering
    \includegraphics[width=\linewidth]
        {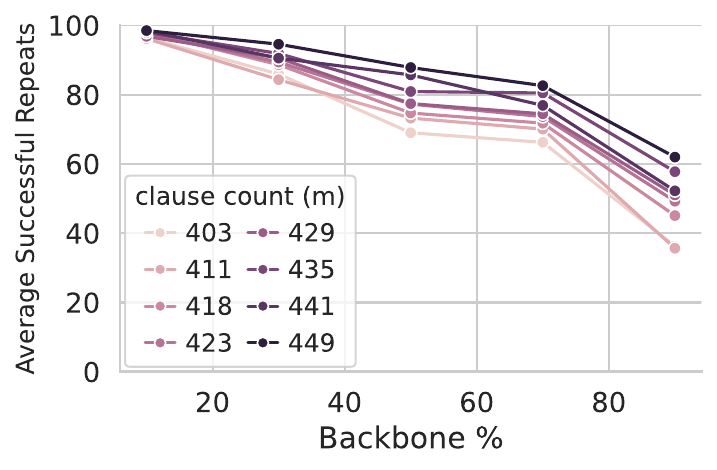}
    \caption{Ising chip}
    \label{fig:backbone_cobi_repeat}
  \end{subfigure}
  \vspace{-0.3cm}
  \caption{Successful repeats as a function of backbone
  density $\beta$ under decomposed execution.}
  \Description{Two line plots of successful repeats versus backbone density for D-Wave Tabu and the Ising chip; success decreases as backbone density increases.}
\label{fig:backbones_repeats}
  \vspace{-0.1cm}
\end{figure}

Figure~\ref{fig:backbones_repeats} indicates that successful repeats decrease monotonically as backbone density increases from 10\% to 90\%, across all clause counts and for both solvers. The consistency across Tabu and the fabricated Ising chip confirms that the degradation is a property of the SAT-to-QUBO formulation, not of a solver or problem class.

\subsection{Preprocessing Impact}
\label{sec:eval_preprocess}

Table~\ref{tab:preprocess} tracks average variable count after each preprocessing stage as a proxy for QUBO size. Actual QUBO size also depends on clause structure\blue{,} since 3SAT clauses introduce ancillary spins whereas 1SAT and 2SAT clauses do not, but variable count captures the main reduction trend.

\begin{table}[t]
\centering
\small
\setlength{\tabcolsep}{3pt}
\caption{Average variable count in the semiprime
CNF after each preprocessing level. Levels~0--5 
constitute the proposed flow; {levels~6--7 are 
ablations included for reference.}}
\label{tab:preprocess}
\fitcolumn{
\begin{tabular}{lcccccc}
\toprule
\textbf{Preprocessing Level} &
\textbf{4b} & \textbf{5b} & \textbf{7b} & 
\textbf{8b} & \textbf{10b} & \textbf{11b} \\
\midrule
\textbf{0} Original        & 21 & 37 & 76 & 94 & 162 & 190 \\
\textbf{1} +Gate Rewriting & 21 & 37 & 76 & 94 & 162 & 190 \\
\textbf{2} +Known Values   & 16 & 29 & 64 & 81 & 145 & 172 \\
\textbf{3} +Binary Relations    & 7  & 7  & 41 & 49 & 90  & 123 \\
\textbf{4} +Replacement\ Propagation   & 0  & 0  & 36 & 45 & 83  & 119 \\
\rowcolor{gray!15} \textbf{5} +Clause Cleaning  & 0  & 0  & 36 & 45 & 83  & 119 \\
\midrule
\textbf{6} +Subsumption    & 0  & 0  & 36 & 45 & 83  & 119 \\
\phantom{\textbf{6}} +Pure Literal Elimination  & 0  & 0  & 35 & 44 & 82  & 118 \\
\textbf{7} +Branching      & 0  & 0  & 8  & 7  & 40  & 73  \\
\bottomrule
\end{tabular}
}
\end{table}

Gate-aware rewriting (level~1) leaves variable count unchanged but reduces downstream QUBO cost by choosing encodings with fewer ancillary spins. Known-value elimination (level~2) produces the first substantial reduction as fixed product bits propagate through the arithmetic circuit. The largest single drop occurs at level~3, where binary relation elimination removes variables that are fully determined by others. Replacement propagation and clause cleaning then remove residual overhead. For 4- and 5-bit instances, preprocessing resolves the problem entirely. For 7--11-bit instances, the residual counts (36, 45, 83, 119) still require decomposition, but the subproblems are substantially smaller.

\subsection{Structure-Aware Decomposition}
\label{sec:eval_decomp}

\begin{figure}[tp]
  \centering
  \begin{subfigure}[b]{0.49\linewidth}
    \centering
    \includegraphics[width=\linewidth]
      {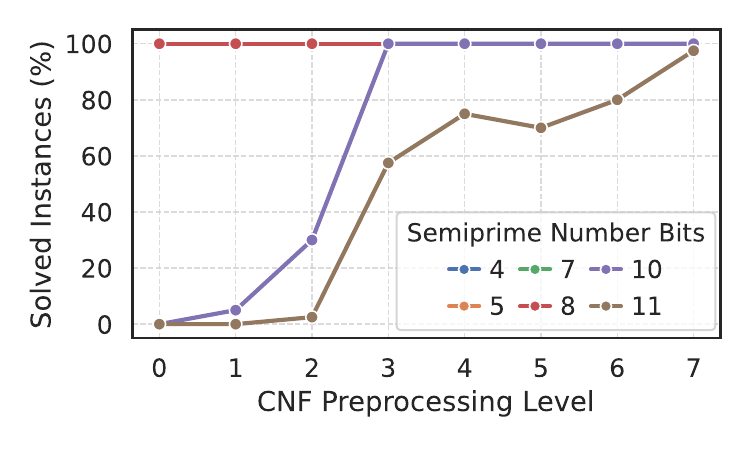}
    \caption{D-Wave Tabu, BFS}
  \end{subfigure}\hfill
  \begin{subfigure}[b]{0.49\linewidth}
    \centering
    \includegraphics[width=\linewidth]
      {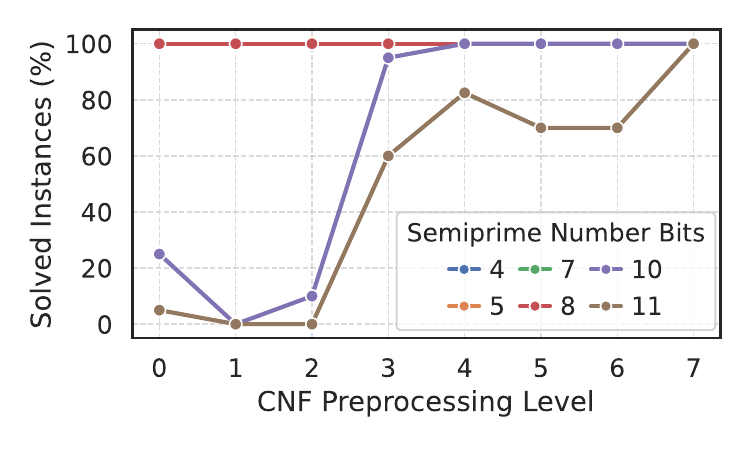}
    \caption{Ising chip, BFS}
  \end{subfigure}
  \begin{subfigure}[b]{0.49\linewidth}
    \centering
    \includegraphics[width=\linewidth]
      {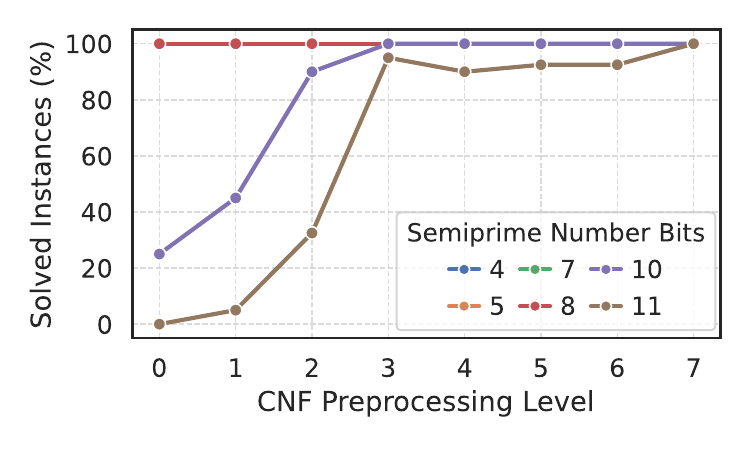}
    \caption{D-Wave Tabu, DFS}
  \end{subfigure}\hfill
  \begin{subfigure}[b]{0.49\linewidth}
    \centering
    \includegraphics[width=\linewidth]
      {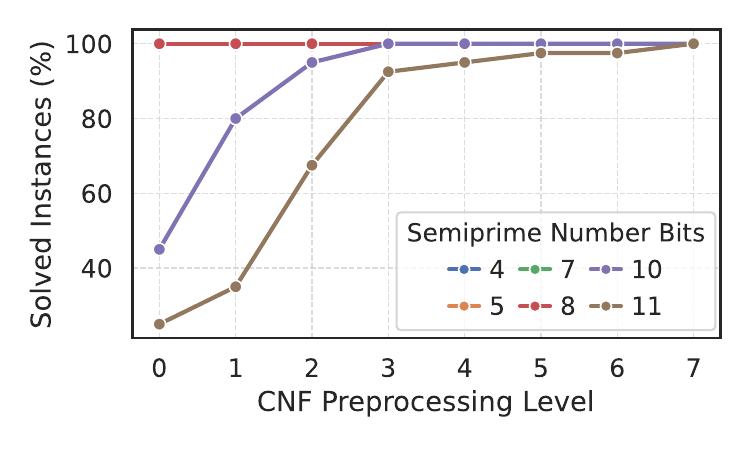}
    \caption{Ising chip, DFS}
  \end{subfigure}
  \caption{\% solved semiprime instances across 
  preprocessing levels under BFS and DFS decomposition.}
  \Description{Four line plots of the percentage of solved semiprime instances versus preprocessing level for D-Wave Tabu and the Ising chip under BFS and DFS decomposition.}
\label{fig:decomposition_solved_instances}
  \vspace{-0.2cm}
\end{figure}

We compare BFS and DFS subproblem selection across all instance sizes and preprocessing levels for both solvers. \blue{Traversal order and initial variable assignments are randomized across independent repeats.} Figure~\ref{fig:decomposition_solved_instances} shows that DFS consistently outperforms BFS across all configurations. Under BFS, solve rates fluctuate as preprocessing progresses---particularly on 10- and 11-bit instances---indicating sensitivity to residual gate-cluster structure. Under DFS, solve rates improve monotonically. The gap is most pronounced on the largest instances, where each subproblem covers only a small fraction of the circuit. The effect is most visible on the fabricated chip, where fixed hardware size makes poor decomposition particularly costly.

\begin{figure}[tp]
  \centering
  \begin{subfigure}[b]{0.49\linewidth}
    \centering
    \includegraphics[width=\linewidth]
      {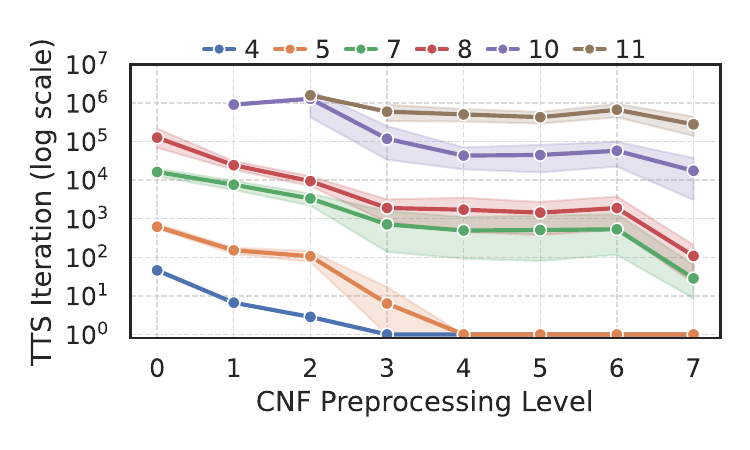}
    \caption{D-Wave Tabu, BFS}
  \end{subfigure}\hfill
  \begin{subfigure}[b]{0.49\linewidth}
    \centering
    \includegraphics[width=\linewidth]
      {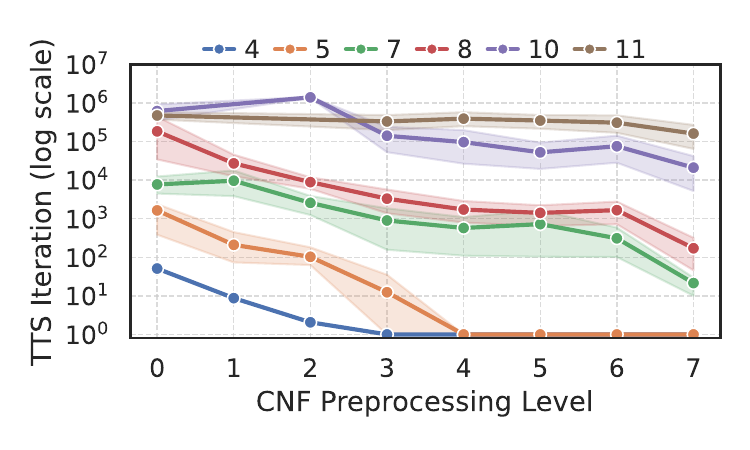}
    \caption{Ising chip, BFS}
  \end{subfigure}
  \begin{subfigure}[b]{0.49\linewidth}
    \centering
    \includegraphics[width=\linewidth]
      {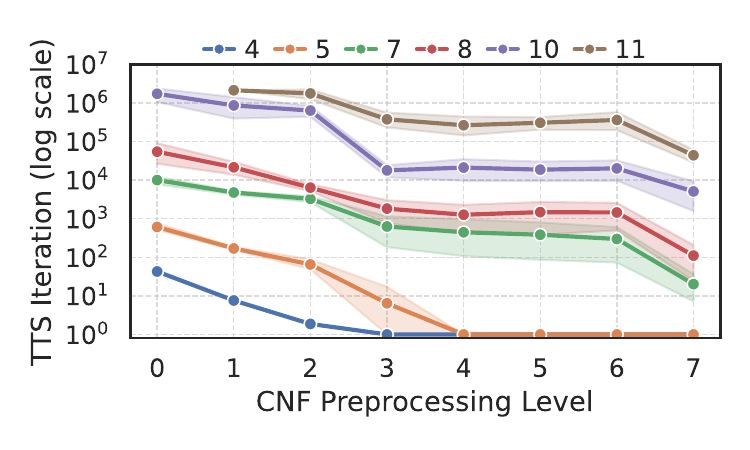}
    \caption{D-Wave Tabu, DFS}
  \end{subfigure}\hfill
  \begin{subfigure}[b]{0.49\linewidth}
    \centering
    \includegraphics[width=\linewidth]
      {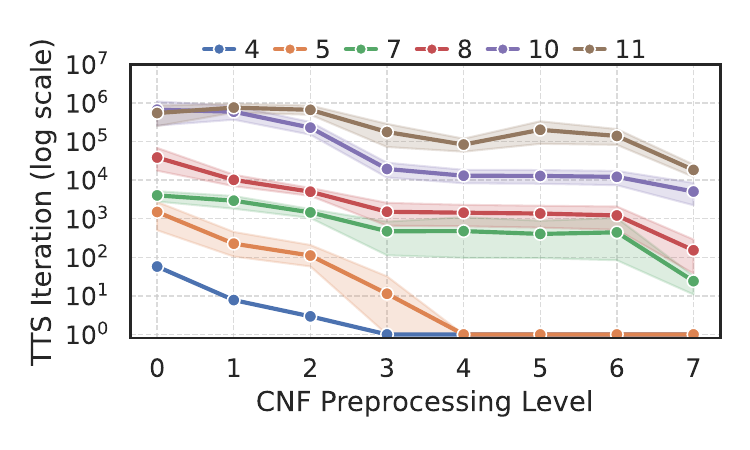}
    \caption{Ising chip, DFS}
  \end{subfigure}
  \caption{TTS \blue{vs.}\ preprocessing level
  under BFS and DFS decomposition.}
  \Description{Four line plots of time to solution versus preprocessing level for D-Wave Tabu and the Ising chip under BFS and DFS decomposition.}
\label{fig:decomposition_tts}
  \vspace{-0.5cm}
\end{figure}

Figure~\ref{fig:decomposition_tts} reports TTS for fully solved groups. DFS yields consistently decreasing TTS as preprocessing progresses; BFS renders TTS fluctuations across levels. Two observations stand out. First, variable count alone does not determine TTS: instances with similar residual sizes can exhibit different convergence behavior depending on the remaining constraint structure. Second, the fabricated Ising chip achieves lower TTS than Tabu at higher preprocessing levels, indicating higher-quality solutions to subproblems.
Together, these results confirm that preprocessing and structure-aware decomposition are complementary: each strengthens the other.

\subsection{Preprocessing Asymmetry: Ising vs.\ CDCL}
\label{sec:eval_cdcl}

\begin{figure}[!tb]
  \centering
  \begin{subfigure}[b]{0.49\linewidth}
    \centering
    \includegraphics[width=\linewidth]
      {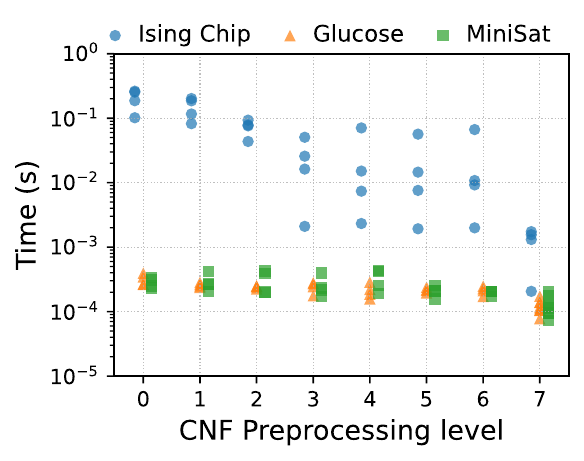}
    \caption{Time vs.\ preprocessing level}
  \end{subfigure}\hfill
  \begin{subfigure}[b]{0.49\linewidth}
    \centering
    \includegraphics[width=\linewidth]
      {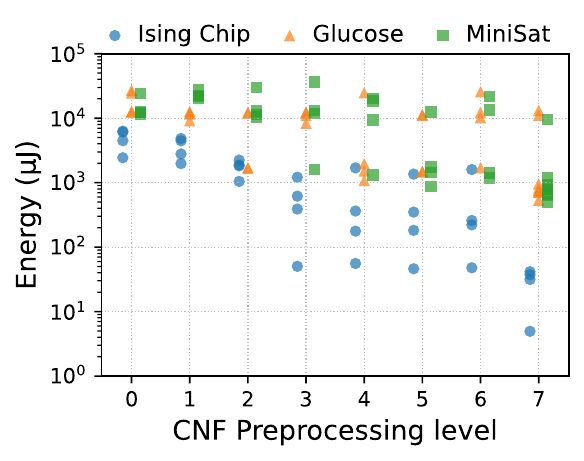}
    \caption{Energy vs.\ preprocessing level}
  \end{subfigure}
  \caption{Time and energy distributions across
  preprocessing levels for representative 7-bit
  semiprime instances.}
  \Description{Two plots of runtime and energy versus preprocessing level for CDCL solvers and the Ising chip on 7-bit semiprime instances.}
\label{fig:cdcl_compare_7bit}
  \vspace{-0.2cm}
\end{figure}

A central implication of the analysis in Section~\ref{sec:limits} is that preprocessing should affect Ising execution far more than symbolic solvers, because the structure it removes is exploited implicitly by CDCL but survives as harmful penalty terms in the Ising formulation. We compare the Ising chip against Glucose and MiniSat on representative 7-bit semiprime instances \blue{(Figure~\ref{fig:cdcl_compare_7bit})}. All solvers receive the same preprocessed CNF at each level. Ising time and energy reflect chip execution only, excluding preprocessing overhead.

Table~\ref{tab:cdcl_ising_gaps_7bit} quantifies the gap. Across all preprocessing levels, the best CDCL runtime varies by less than $3\times$; measured Ising runtime varies by more than $150\times$. At level~0, CDCL is $434\times$ faster, yet the Ising chip already shows \blue{a} $4.8\times$ energy advantage\blue{; by} level~7, the runtime gap narrows to $2.8\times$ and the energy advantage exceeds $100\times$.

\begin{table}[!tb]
\centering
\small
\caption{Best-observed runtime gap vs. energy advantage
on representative 7-bit semiprime instances. 
Runtime gap\,=\,best Ising time\,/\,best CDCL 
time; energy advantage\,=\,best CDCL 
energy\,/\,best Ising energy.}
\label{tab:cdcl_ising_gaps_7bit}
\setlength{\tabcolsep}{6pt}
\begin{tabular}{ccc}
\toprule
\textbf{Level} & 
\textbf{Runtime Gap} $\downarrow$ & 
\textbf{Energy Advantage} $\uparrow$ \\
\midrule
0 & 433.89$\times$ & 4.76$\times$  \\
1 & 398.45$\times$ & 4.63$\times$  \\
2 & 219.86$\times$ & 1.59$\times$  \\
3 & 12.20$\times$  & 31.94$\times$ \\
4 & 14.73$\times$  & 19.48$\times$ \\
5 & 12.32$\times$  & 18.94$\times$ \\
6 & 11.58$\times$  & 24.60$\times$ \\
7 & 2.82$\times$   & 101.50$\times$ \\
\bottomrule
\end{tabular}
\end{table}

The exact same preprocessing that changes CDCL runtime by less than $3\times$ changes Ising runtime by over $150\times$. This indicates that the bottleneck is not raw Ising machine capability, but the form in which structured SAT is exposed to the Ising machine.

\subsection{Runtime Tradeoff of Preprocessing}
\label{sec:eval_runtime}

\begin{figure}[!tb]
  \centering
  \begin{subfigure}[b]{0.49\linewidth}
    \centering
    \includegraphics[width=\linewidth]
      {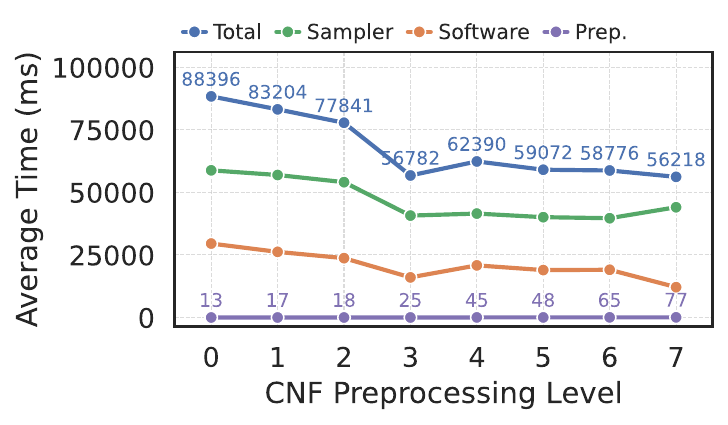}
    \caption{D-Wave Tabu}
  \end{subfigure}\hfill
  \begin{subfigure}[b]{0.49\linewidth}
    \centering
    \includegraphics[width=\linewidth]
      {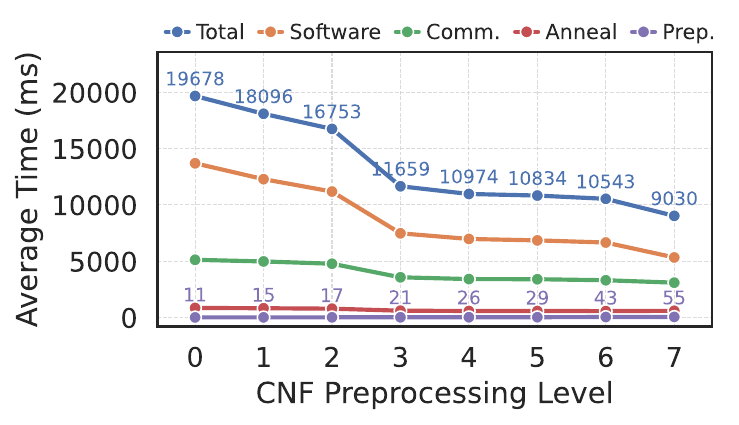}
    \caption{Ising chip}
  \end{subfigure}
  \caption{Preprocessing overhead versus solver time across preprocessing levels\blue{, averaged over all semiprime instances and runs under DFS decomposition; solver phases are per-iteration costs scaled by the average iteration count. Software: decomposition, QUBO conversion, and mapping; Comm.: PCIe communication; Anneal: chip anneal; Sampler: Tabu calls; Total: their sum; Prep.: preprocessing}.}
  \Description{Two line plots of preprocessing time and per-phase solver time versus preprocessing level for D-Wave Tabu and the Ising chip.}
\label{fig:runtime_tradeoff}
  \vspace{-0.1cm}
\end{figure}

Figure~\ref{fig:runtime_tradeoff} breaks down total runtime into preprocessing and solver execution\blue{, averaged over all semiprime instances and runs}. Solver execution includes all software overheads beyond preprocessing (decomposition, QUBO conversion, hardware mapping, and chip time). The entire flow is implemented in unoptimized Python~3.8.10.

Known-value elimination and binary-relation removal produce the largest reduction in solver time at minimal preprocessing cost \blue{(17--25\,ms)}. Subsequent stages through clause cleaning provide modest additional gains \blue{on the chip}. Subsumption and pure literal elimination incur longer preprocessing times with little solver improvement. Branching increases the number of timeout-limited runs, raising \blue{chip time} rather than reducing it. Beyond level~5, preprocessing cost grows faster than \blue{chip-time} benefit. Because the Python orchestration dominates end-to-end runtime \blue{on the chip}, a compiled implementation~\cite{pereira2021ranking} or hardware-offloaded decomposition~\cite{yin2026fpga} would substantially reduce this overhead.

\subsection{End-to-End Results}
\label{sec:eval_e2e}

Table~\ref{tab:e2e_results} summarizes end-to-end results for the largest benchmark groups on the fabricated Ising chip with DFS decomposition. Instances below 10~bits reach full solvability at or before level~3 and are omitted, since their success reflects preprocessing alone rather than the quality of the decomposition strategy.

\begin{table}[!tb]
\centering
\small
\caption{Solved instance percentage (\%) at each
preprocessing level, fabricated Ising chip with 
DFS decomposition. The proposed flow is level~5; 
levels~6--7 are ablations.}
\label{tab:e2e_results}
\setlength{\tabcolsep}{4pt}
\fitcolumn{
\begin{tabular}{lccccc>{\columncolor{gray!15}}ccc}
\toprule
\textbf{Bits} & 
\textbf{L0} & \textbf{L1} & \textbf{L2} & 
\textbf{L3} & \textbf{L4} & \textbf{L5} & 
\textbf{L6} & \textbf{L7} \\
\midrule
10 & 45.0 & 80.0 & 95.0 & 
   \textbf{100.0} & 100.0 & 100.0 & 100.0 & 100.0 \\
11 & 25.0 & 35.0 & 67.5 & 
   92.5 & 95.0 & \textbf{97.5} & 97.5 & 100.0 \\
\bottomrule
\end{tabular}
}
\vspace{-0.3cm}
\end{table}

For 10-bit instances, solve rate reaches 100\% at level~3. For 11-bit instances, the proposed flow (level~5) reaches 97.5\% (39/40). Full solvability appears only at level~7, which incurs disproportionate preprocessing cost (Section~\ref{sec:eval_runtime}). The dominant step in both groups is level~3, which raises 11-bit solve rate from 67.5\% to 92.5\%.

These results show that the 8-bit ceiling of the unpreprocessed baseline is a consequence of unremoved tight constraints rather than a fundamental hardware limit. On the same Ising chip, without hardware changes, our proposed flow extends the solvable range from 8-bit (94 variables) to 11-bit (190 variables) --- more than doubling the largest solvable problem size.

\blue{\noindent\textbf{Beyond multiplier circuits:} ATPG and equivalence-checking instances generated from 74-series and ISCAS-85 circuits~\cite{efe2026benchmarks} show the same trend---the smaller 74-series circuits are always solvable and only accelerate, while the ISCAS-85 controller c432 becomes fully solvable at level~3 on both the chip and D-Wave Tabu; full results are provided as supplementary material.}

\section{Conclusion}
\label{sec:conclusion}
This paper identifies tight constraints as the primary bottleneck
for Ising-based solvers on circuit-derived SAT. When mapped to
a QUBO objective, these constraints distort the energy landscape
without providing the logical guidance necessary for search. Under
decomposition, they facilitate the propagation of erroneous
assignments (from frozen, stationary variables), while on resource-constrained
hardware, they consume a physical budget that should
be dedicated to the thorough exploration of the search space.
Our proposed framework mitigates these effects through a constraint-aware
preprocessing flow that excises logical redundancies before
they reach the hardware, coupled with a DFS-based decomposition
that preserves the signal-flow context discarded by alternative
neighborhood-centric traversals. Remarkably, preprocessing that
yields less than a 3$\times$ performance delta for classical (CDCL) solvers
results in an over 150$\times$ improvement for Ising machines. This
disparity demonstrates that the actual bottleneck is not the raw machine
capability, but rather the representation of structured SAT instances.
Evaluated on a fabricated 45-spin CMOS Ising platform, our flow
effectively doubles the solvable problem size (from 8-bit to 11-bit
semiprime instances) without hardware modification. These results
suggest that future Ising architectures would benefit from native
support for hard constraints and deterministic side information,
rather than encoding all structure solely as QUBO penalty terms.


\bibliography{ref}

@article{mosca2022factoring,
  title="{Factoring semi-primes with (quantum) SAT-solvers}",
  author={Mosca, Michele and Verschoor, Sebastian R.},
  journal={Scientific Reports},
  volume={12},
  number={1},
  pages={7982},
  doi={10.1038/s41598-022-11687-7},
  year={2022},
  publisher={Nature Publishing Group UK London},
}

@incollection{tseitin1983complexity,
  title={On the Complexity of Derivation in Propositional Calculus},
  doi={10.1007/978-3-642-81955-1_28},
  author={Tseitin, Grigori S.},
  booktitle={Automation of reasoning: 2: Classical papers on computational logic 1967--1970},
  pages={466--483},
  year={1983},
  publisher={Springer}
}

@incollection{hoos2000satlib,
  author    = {Hoos, Holger H. and St{\"u}tzle, Thomas},
  title     = {SATLIB: An Online Resource for Research on SAT},
  booktitle = {SAT 2000: Highlights of Satisfiability Research in the Year 2000},
  editor    = {Gent, Ian P. and van Maaren, Hans and Walsh, Toby},
  pages     = {283--292},
  publisher = {IOS Press},
  year      = {2000}
}

@article{jun2023hubo,
  author    = {Kyungtaek Jun and Hyunju Lee},
  title     = {{HUBO} and {QUBO} Models for Prime Factorization},
  journal   = {Scientific Reports},
  volume    = {13},
  number    = {1},
  pages     = {10080},
  year      = {2023},
  doi       = {10.1038/s41598-023-36813-x},
  publisher = {Nature Publishing Group UK London}
}

@article{jiang2018quantum,
  author    = {Shuxian Jiang and Keith A. Britt and Alexander J. McCaskey and Travis S. Humble and Sabre Kais},
  title     = {Quantum Annealing for Prime Factorization},
  journal   = {Scientific Reports},
  volume    = {8},
  number    = {1},
  pages     = {17667},
  year      = {2018},
  doi       = {10.1038/s41598-018-36058-z},
  publisher = {Nature Publishing Group UK London}
}

@techreport{boost2017partitioning,
  author      = {Michael Booth and Steven P. Reinhardt and Aidan Roy},
  title       = {Partitioning Optimization Problems for Hybrid Classical/Quantum Execution},
  institution = {D-Wave Systems Inc.},
  address     = {Burnaby, Canada},
  number      = {14-1006A-A},
  year        = {2017},
  type        = {Technical Report}
}

@inproceedings{biere1999symbolic,
  author    = {Armin Biere and Alessandro Cimatti and Edmund M. Clarke and Masahiro Fujita and Yunshan Zhu},
  doi       = {10.1145/309847.309942},
  title     = "{Symbolic model checking using SAT procedures instead of BDDs}",
  booktitle = {Proceedings of the 36th Annual ACM/IEEE Design Automation Conference (DAC)},
  pages     = {317--320},
  year      = {1999}
}

@inproceedings{cheeseman1991really,
  author    = {Peter C. Cheeseman and Bob Kanefsky and William M. Taylor},
  title     = "{Where the Really Hard Problems Are}",
  booktitle = {Proceedings of the 12th International Joint Conference on Artificial Intelligence (IJCAI)},
  volume    = {1},
  pages     = {331--337},
  year      = {1991}
}

@inproceedings{mitchell1992hard,
  author       = {David Mitchell and Bart Selman and Hector Levesque},
  title        = "{Hard and Easy Distributions of {SAT} Problems}",
  booktitle    = {Proceedings of the 10th National Conference on Artificial Intelligence (AAAI)},
  pages        = {459--465},
  year         = {1992},
  publisher    = {AAAI Press}
}

@article{ansotegui2019community,
  title   = {Community Structure in Industrial {SAT} Instances},
  author  = {Ans{\'o}tegui, Carlos and Bonet, Maria Luisa and Gir{\'a}ldez-Cru, Jes{\'u}s and Levy, Jordi and Simon, Laurent},
  journal = {Journal of Artificial Intelligence Research},
  volume  = {66},
  pages   = {443--472},
  year    = {2019},
  doi     = {10.1613/jair.1.11741}
}

@incollection{gomes2008satisfiability,
  title     = {Satisfiability Solvers},
  author    = {Gomes, Carla P. and Kautz, Henry and Sabharwal, Ashish and Selman, Bart},
  booktitle = {Handbook of Knowledge Representation},
  editor    = {van Harmelen, Frank and Lifschitz, Vladimir and Porter, Bruce},
  series    = {Foundations of Artificial Intelligence},
  volume    = {3},
  pages     = {89--134},
  year      = {2008},
  doi       = {10.1016/S1574-6526(07)03002-7},
  publisher = {Elsevier, Amsterdam, The Netherlands}
}

@article{cilasun2025cobi,
  title     = {A Coupled-Oscillator-Based Ising Chip for Combinatorial Optimization},
  author    = {C{\i}lasun, H{\"u}srev and Moy, William and Zeng, Ziqing and Islam, Tahmida and Lo, Hao and Vanasse, Alex and Tan, Megan and Anees, Mohammad and S, Ramprasath and Kumar, Abhimanyu and Sapatnekar, Sachin S. and Kim, Chris H. and Karpuzcu, Ulya R.},
  journal   = {Nature Electronics},
  volume    = {8},
  number    = {6},
  pages     = {537--546},
  year      = {2025},
  doi       = {10.1038/s41928-025-01393-3},
  publisher = {Springer Nature}
}

@article{cilasun2024sat,
  title     = {3SAT on an All-to-All-Connected CMOS Ising Solver Chip},
  author    = {C{\i}lasun, H{\"u}srev and Zeng, Ziqing and S, Ramprasath and Kumar, Abhimanyu and Lo, Hao and Cho, William and Moy, William and Kim, Chris H. and Karpuzcu, Ulya R. and Sapatnekar, Sachin S.},
  journal   = {Scientific Reports},
  volume    = {14},
  number    = {1},
  pages     = {10757},
  year      = {2024},
  doi       = {10.1038/s41598-024-60316-y},
  publisher = {Nature Publishing Group UK London}
}

@article{johnson2011quantum,
  title     = {Quantum annealing with manufactured spins},
  author    = {Johnson, M. W. and Amin, M. H. S. and Gildert, S. and Lanting, T. and Hamze, F. and Dickson, N. and Harris, R. and Berkley, A. J. and Johansson, J. and Bunyk, P. and Chapple, E. M. and Enderud, C. and Hilton, J. P. and Karimi, K. and Ladizinsky, E. and Ladizinsky, N. and Oh, T. and Perminov, I. and Rich, C. and Thom, M. C. and Tolkacheva, E. and Truncik, C. J. S. and Uchaikin, S. and Wang, J. and Wilson, B. and Rose, G.},
  journal   = {Nature},
  volume    = {473},
  number    = {7346},
  pages     = {194--198},
  year      = {2011},
  doi       = {10.1038/nature10012}
}

@article{inagaki2016coherent,
  title     = {A coherent Ising machine for 2000-node optimization problems},
  author    = {Inagaki, Takahiro and Haribara, Yoshitaka and Igarashi, Koji and Sonobe, Tomohiro and Tamate, Shuhei and Honjo, Toshimori and Marandi, Alireza and McMahon, Peter L. and Umeki, Takeshi and Enbutsu, Koji and Tadanaga, Osamu and Takenouchi, Hirokazu and Aihara, Kazuyuki and Kawarabayashi, Ken-ichi and Inoue, Kyo and Utsunomiya, Shoko and Takesue, Hiroki},
  journal   = {Science},
  volume    = {354},
  number    = {6312},
  pages     = {603--606},
  year      = {2016},
  doi       = {10.1126/science.aah4243}
}

@article{lucas2014ising,
  title     = {{Ising} Formulations of Many {NP} Problems},
  author    = {Lucas, Andrew},
  journal   = {Frontiers in Physics},
  volume    = {2},
  pages     = {5},
  year      = {2014},
  doi       = {10.3389/fphy.2014.00005},
  publisher = {Frontiers Media S.A., Lausanne, Switzerland}
}

@article{glover2018tutorial,
  title     = {A Tutorial on Formulating and Using {QUBO} Models},
  author    = {Glover, Fred and Kochenberger, Gary and Du, Yu},
  journal   = {arXiv preprint arXiv:1811.11538},
  year      = {2018},
  note      = {Available at \url{https://arxiv.org/abs/1811.11538}}
}

@article{chancellor2016direct,
  title     = {A Direct Mapping of Max {k}-{SAT} and High Order Parity Checks to a Chimera Graph},
  author    = {Chancellor, Nicholas and Zohren, Stefan and Warburton, Paul A. and Benjamin, Simon C. and Roberts, Stephen},
  journal   = {Scientific Reports},
  volume    = {6},
  number    = {1},
  pages     = {37107},
  year      = {2016},
  doi       = {10.1038/srep37107},
  publisher = {Nature Publishing Group, London, United Kingdom}
}

@misc{dwave-tabu,
  author       = {{D-Wave Systems Inc.}},
  title        = {dwave-tabu: An Optimized Tabu Search Solver for QUBO Problems},
  howpublished = {\url{https://pypi.org/project/dwave-tabu/}},
  year         = {2022},
  note         = {Version 0.5.0}
}

@article{ronnow2014defining,
  title        = {Defining and Detecting Quantum Speedup},
  author       = {R{\o}nnow, Troels F. and Wang, Zhihui and Job, Joshua and Boixo, Sergio and Isakov, Sergei V. and Wecker, David and Martinis, John M. and Lidar, Daniel A. and Troyer, Matthias},
  journal      = {Science},
  volume       = {345},
  number       = {6195},
  pages        = {420--424},
  year         = {2014},
  doi          = {10.1126/science.1252319}
}

@misc{quicc_sat_datasets,
  author       = {Regli, William and Mossi, Gianni and Hajiaghayi, Mohammad Taghi and Munoz Bauza, Humberto and Whitehouse, Ian Joseph and Banihashem, Kiarash and Jabbarzade, Peyman and Paul, Taylor Henry},
  title        = {SAT Benchmarks to Assess Quantum-Inspired Solvers},
  howpublished = {GitHub repository},
  year         = {2025},
  url          = {https://github.com/UMD-ARLIS/QuICC-SAT-Datasets/tree/main},
  note         = {UMD-ARLIS / QuICC-SAT-Datasets}
}

@article{larrabee1992test,
  author    = {Tracy Larrabee},
  title     = {Test Pattern Generation Using {Boolean} Satisfiability},
  journal   = {IEEE Transactions on Computer-Aided Design of Integrated Circuits and Systems},
  volume    = {11},
  number    = {1},
  pages     = {4--15},
  year      = {1992},
  doi       = {10.1109/43.108614}
}

@article{ahmed2021probabilistic,
  author    = {Ibrahim Ahmed and Po-Wei Chiu and William Moy and Chris H. Kim},
  title     = {A Probabilistic Compute Fabric Based on Coupled Ring Oscillators for Solving Combinatorial Optimization Problems},
  journal   = {IEEE Journal of Solid-State Circuits},
  volume    = {56},
  number    = {9},
  pages     = {2870--2880},
  year      = {2021},
  doi       = {10.1109/JSSC.2021.3062821}
}

@article{kuehlmann2002robust,
  author    = {Andreas Kuehlmann and Viresh Paruthi and Florian Krohm and Malay K. Ganai},
  title     = {Robust {Boolean} Reasoning for Equivalence Checking and Functional Property Verification},
  journal   = {IEEE Transactions on Computer-Aided Design of Integrated Circuits and Systems},
  volume    = {21},
  number    = {12},
  pages     = {1377--1394},
  year      = {2002},
  doi       = {10.1109/TCAD.2002.804386}
}

@article{zhang2024skisat,
  author    = {Ahmet Yusuf Salim and Bart Selman and Henry Kautz and Zeljko Ignjatovic and Sel\c{c}uk K\"{o}se},
  title     = {{SKI-SAT}: A {CMOS}-Compatible Hardware for Solving {SAT} Problems},
  journal   = {IEEE Transactions on Circuits and Systems I: Regular Papers},
  volume    = {72},
  number    = {10},
  pages     = {5926--5937},
  year      = {2025},
  doi       = {10.1109/TCSI.2025.3554145}
}

@inproceedings{tan2023hyqsat,
  author    = {Siwei Tan and Mingqian Yu and Andre Python and Yongheng Shang and Tingting Li and Liqiang Lu and Jianwei Yin},
  title     = {{HyQSAT}: A Hybrid Approach for 3-{SAT} Problems by Integrating Quantum Annealer with {CDCL}},
  booktitle = {Proc. IEEE International Symposium on High-Performance Computer Architecture (HPCA)},
  pages     = {731--744},
  year      = {2023},
  doi       = {10.1109/HPCA56546.2023.10071022}
}

@inproceedings{nusslein2023solving,
  author    = {Jonas N{\"u}{\ss}lein and Sebastian Zielinski and Thomas Gabor and Claudia Linnhoff-Popien and Sebastian Feld},
  title     = {Solving (Max) 3-{SAT} via Quadratic Unconstrained Binary Optimization},
  booktitle = {Proc. International Conference on Computational Science (ICCS)},
  series    = {LNCS},
  volume    = {14077},
  pages     = {34--47},
  year      = {2023},
  doi       = {10.1007/978-3-031-36030-5_3}
}

@inproceedings{sikhakollu2024circuit,
  author    = {Venkata Pavan Sumanth Sikhakollu and Shreesha Sreedhara and Rajit Manohar and Alan Mishchenko and Jaijeet Roychowdhury},
  title     = {High Quality Circuit-Based 3-{SAT} Mappings for Oscillator {Ising} Machines},
  booktitle = {Proc. International Conference on Unconventional Computation and Natural Computation (UCNC)},
  series    = {LNCS},
  volume    = {14776},
  pages     = {269--285},
  year      = {2024},
  doi       = {10.1007/978-3-031-63742-1_19}
}

@article{jagielski2023fpim,
  author    = {Thomas Jagielski and Rajit Manohar and Jaijeet Roychowdhury},
  title     = {{FPIM}: Field-Programmable {Ising} Machines for Solving {SAT}},
  journal   = {arXiv preprint arXiv:2306.01569},
  year      = {2023}
}

@article{bybee2023higher,
  author    = {Connor Bybee and Denis Kleyko and Dmitri E. Nikonov and Amir Khosrowshahi and Bruno A. Olshausen and Friedrich T. Sommer},
  title     = {Efficient Optimization with Higher-Order {Ising} Machines},
  journal   = {Nature Communications},
  volume    = {14},
  pages     = {6033},
  year      = {2023},
  doi       = {10.1038/s41467-023-41214-9}
}

@article{ding2024factorization,
  author    = {Jingwen Ding and Giuseppe Spallitta and Roberto Sebastiani},
  title     = {Effective Prime Factorization via Quantum Annealing by Modular Locally-Structured Embedding},
  journal   = {Scientific Reports},
  volume    = {14},
  number    = {1},
  pages     = {3518},
  year      = {2024},
  doi       = {10.1038/s41598-024-53708-7}
}

@inproceedings{pysat,
  author    = {Alexey Ignatiev and
               Antonio Morgado and
               Joao Marques-Silva},
  title     = {{PySAT}: {A} {Python} Toolkit for
               Prototyping with {SAT} Oracles},
  booktitle = {Proceedings of the 21st International
               Conference on Theory and Applications
               of Satisfiability Testing (SAT)},
  pages     = {428--437},
  year      = {2018},
  doi       = {10.1007/978-3-319-94144-8_26}
}

@article{pereira2021ranking,
  author    = {Rui Pereira and Marco Couto and Francisco Ribeiro and Rui Rua and J{\'a}come Cunha and Jo{\~a}o Paulo Fernandes and Jo{\~a}o Saraiva},
  title     = {Ranking Programming Languages by Energy Efficiency},
  journal   = {Science of Computer Programming},
  volume    = {205},
  pages     = {102609},
  year      = {2021},
  doi       = {10.1016/j.scico.2021.102609}
}

@inproceedings{efe2026satic,
  doi          = {10.1109/ISCA66397.2026.00165},
  title        = {{SATIC}: An Optimizing {Ising} Compiler for {SAT}(isfiability)},
  author       = {Efe, Ahmet and C{\i}lasun, H{\"u}srev and Kumar, Abhimanyu and Prova, Nafisa S. and Zeng, Ziqing and Islam, Tahmida and Yin, Ruihong and Li, Chaohui and Kreye, Peter and Kim, Chris H. and Sapatnekar, Sachin S. and Karpuzcu, Ulya R.},
  booktitle    = {2026 ACM/IEEE 53rd Annual International Symposium on Computer Architecture (ISCA)},
  pages        = {2349--2363},
  year         = {2026},
  organization = {IEEE}
}

@misc{efe2026benchmarks,
  author       = {Ahmet Efe and H{\"u}srev C{\i}lasun and Abhimanyu Kumar and Nafisa S. Prova and Ziqing Zeng and Tahmida Islam and Ruihong Yin and Chaohui Li and Peter Kreye and Chris H. Kim and Sachin S. Sapatnekar and Ulya R. Karpuzcu},
  title        = {Circuit-Derived {SAT} Benchmarks: {ATPG} and Equivalence Checking},
  howpublished = {GitHub repository},
  year         = {2026},
  url          = {https://github.com/ahmetefe3423/circuit-derived-sat-benchmarks}
}

@article{yin2026fpga,
  title     = {Decomposing Large-Scale {I}sing Problems on {FPGA}s: A Hybrid Hardware Approach},
  author    = {Yin, Ruihong and Zheng, Yue and Li, Chaohui and Efe, Ahmet and Kumar, Abhimanyu and Zeng, Ziqing and Karpuzcu, Ulya R. and Sapatnekar, Sachin S. and Kim, Chris H.},
  journal   = {arXiv preprint arXiv:2602.15985},
  year      = {2026},
  note      = {Available at \url{https://arxiv.org/abs/2602.15985}}
}

\end{document}